\documentclass[prl,preprint,amsmath,amssymb]{revtex4}
\usepackage{graphicx}
\usepackage[justification=centering]{caption}

\def\ra{\rangle}
\def\la{\langle}

\begin{document}

\title{Triangle-like inequalities related to coherence and entanglement negativity}

\author{Zhi-Xiang Jin$^{1}$}
\author{Xianqing Li-Jost$^{2,3}$}
\author{Shao-Ming Fei$^{1,2}$}

\affiliation{$^1$School of Mathematical Sciences, Capital Normal University,
Beijing 100048, China\\
$^2$Max-Planck-Institute for Mathematics in the Sciences, 04103 Leipzig, Germany\\
$^3$School of Mathematics and Statistics, Hainan Normal University, Haikou 571158, China}
\bigskip

\begin{abstract}

Quantum coherence and entanglement are two key features in quantum mechanics and play important roles in quantum information processing and quantum computation. We provide a general triangle-like inequality satisfied by the $l_1$-norm measure of coherence for convex combination of arbitrary $n$ pure states of a quantum state $\rho$. Furthermore, we present triangle-like inequality for the convex-roof extended negativity for any states of rank 2, which gives a positive answer to a conjecture raised in [Phys. Rev. A 96, 062308 (2017)]. Detailed examples are given to illustrate the relations characterized by the triangle-like inequalities.
\end{abstract}

\maketitle

\section{INTRODUCTION}
Quantum coherence and entanglement are the two key features of quantum world. While quantum coherence is defined for single systems, quantum entanglement is adopted to describe the correlations in bipartite or multipartite systems. Recent developments in understanding of quantum coherence have come from the burgeoning field of quantum information science. Just like entanglement, coherence can be treated as a physical resource and has been widely studied.  One important pillar of the field is the study on quantification of coherence. Since the seminal work \cite{tmm} on defining a good coherence measure in terms of the resource theory, quantum coherence has been widely studied and applied to many quantum information processing  \cite{tmm,spm,dg,ctmm,chs,jbdv,auhm,csb,ad,easm,em,eg,ir,irp,yxlc,ula,aer,mmtc,cmst,bukf}.

The relative entropy and $l_1$-norm coherence measures are two well-known measures of coherence, especially concerning the strong monotonicity property and the closed expressions.
In fact, different quantifications of coherence can greatly enrich our understanding of coherence. In particular, the distillable coherence \cite{ad,yzcm}, the coherence of formation \cite{ad,yzcm}, the robustness of coherence \cite{ctmm}, the coherence measures based on entanglement \cite{auhm}, max-relative entropy of coherence \cite{bukf}, and the coherence concurrence \cite{gt,dbq} have been proposed and investigated.
For instance, the relative entropy coherence can be understood as the optimal rate for distilling a maximally coherent state from given states \cite{ad}. The max-relative entropy of coherence can exactly characterize the subchannel discrimination problems such that the coherent state allows for a higher probability of successfully discriminating subchannels than that of all incoherent states \cite{bukf}.
The robustness of coherence quantifies the advantage enabled by a quantum state in a phase discrimination task \cite{ctmm}. In addition,
the relations between coherence and path information \cite{bosp,bbch,bukf1}, the distribution of quantum coherence in multipartite systems \cite{cmst}, the complementarity between coherence and mixedness \cite{ch,sbdp} have also been studied.

In \cite{dy}, the authors show that if a rank-2 state $\rho$ can be expressed as a convex combination of two pure states,  i.e., $\rho=p_1|\psi_1\ra\la\psi_1|+p_2|\psi_2\ra\la\psi_2|$, a triangle inequality can be established, $|E(\sqrt{p_1}|\psi_1\ra)-E(\sqrt{p_2}|\psi_2\ra)|\leq E(\rho)\leq E(\sqrt{p_1}|\psi_1\ra)+E(\sqrt{p_2}|\psi_2\ra)$, where $E$ can be either the measures of coherence or the entanglement concurrence.
In this paper, we give a revise to the inequalities to make it a suitable measure. We provide detailed proofs of triangle-like inequalities in coherence measures and entanglement negativity. A general triangle-like inequality for any convex combination of $n$ states $\rho$ based on the $l_1$ norm coherence measure is given. Furthermore, we provide entanglement negativity satisfied the triangle-like inequality for any states of rank 2, which gives a positive answer to a conjecture of the convex-roof extended negativity satisfied the triangle-like inequality \cite{dy}. At last, we give an example of entanglement negativity in a two-qubit system.

\section{triangle-like inequalities for measures of coherence}
A widely used measure of coherence is the distance-based measure \cite{tmm}, defined by the minimal
distance between a given state and the set of incoherent quantum states $\mathcal{I}$. The incoherent states are diagonal ones in the reference basis $\{|i\ra\}$ of a $d$-dimensional Hilbert space, $\delta=\sum_{i=1}^d\delta_i|i\ra\la i|$.
A measure of coherence for a state $\rho$ can be defined by
\begin{eqnarray}\label{}
C_D(\rho)=\min_{\delta\in \mathcal{I}}D(\rho,\delta),
\end{eqnarray}
where $D(\rho,\delta)$ denotes certain distance measures of quantum states.

Typical distance-based measures of coherence are the relative entropy, the $l_1$-norm and the trace norm \cite{tmm}. The $l_1$-norm measure of coherence for a state $\rho$ is given by
\begin{eqnarray}\label{l1}
C_{l_1}(\rho)=\min_{\delta\in \mathcal{I}}||\rho-\delta||_{l_1}=\sum_{i\ne j}|\rho_{ij}|,
\end{eqnarray}
which is equal to sum of the absolute values of all off-diagonal elements of $\rho$.

For mixed states, the convex-roof $l_1$-norm is adopted as a different measure of coherence \cite{gt}.
The convex-roof $l_1$-norm of a mixed state $\rho$ is given by
\begin{eqnarray}
\tilde{C_{l_1}}(\rho)=\min_{\{p_i,|\psi_i\ra\}}\sum_ip_iC_{l_1}(|\psi_i\ra),
\end{eqnarray}
where the minimization is taken over all pure state decompositions of $\rho=\sum_i p_i|\psi_i\ra\la \psi_i|,~\sum_ip_i=1$, $C_{l_1}(|\psi_i\ra)$ is the $l_1$-norm of the state $|\psi_i\ra\la\psi_i|$.

We begin with a general triangle-like inequality based on the $l_1$ norm measure of coherence.

{[\bf Theorem 1]}. If a state $\rho$ can be expressed as a convex
combination of $n~(n\geq2)$ states $\rho=\sum_{i=1}^np_i\rho_i$, $C_{l_1}(\rho)$ satisfies the following triangle-like inequality,
\begin{eqnarray}\label{th1}
   \frac{1}{n}\sum_{k=1}^n\bigg|G_k^{(n-1)}-p_kC_{l_1}(\rho_k)\bigg|\leq C_{l_1}(\rho)\leq\sum_i p_iC_{l_1}(\rho_i),
  \end{eqnarray}
where $G_k^{(n-1)}$, $k=1,2,\cdots,n$, are the low bounds of $\sum_{i\neq k}^{n}p_i C_{l_1}\left(\frac{\sum_{j\neq k}^{n}p_j\rho_j}{\sum_{t\neq k}^{n}p_t}\right)$.

{\sf [Proof].} First we consider the case of $n=2$, i.e. $\rho=p_1\rho_1+p_2\rho_2$. Then we have
\begin{eqnarray}\label{pfth11}
C_{l_1}(\rho)&&=C_{l_1}(p_1\rho_1+p_2\rho_2)\nonumber\\
&&=\sum_{i\ne j}|p_1{\rho_1}_{ij}+p_2{\rho_2}_{ij}|\nonumber\\
&&\geq \Big|\sum_{i\ne j}p_1|{\rho_1}_{ij}|-\sum_{i\ne j}p_2|{\rho_2}_{ij}|\Big|\nonumber\\
&&= \Big|p_1C_{l_1}(\rho_1)-p_2C_{l_1}(\rho_2)\Big|,
\end{eqnarray}
where $G_1^{(1)}=p_2C_{l_1}(\rho_2)$, $G_2^{(1)}=p_1C_{l_1}(\rho_1)$. Hence (\ref{pfth11}) is a special case of the left hand of (\ref{th1}).

Next, we consider the case of $n=3$, $\rho=\sum_{i=1}^3p_i\rho_i$. From (\ref{pfth11}), we get

\begin{eqnarray}\label{pfth12}
C_{l_1}(\rho)&&\geq \Bigg|\sum_{j\neq i}^{3}p_jC_{l_1}\left(\frac{\sum_{t\neq i}^{3}p_t\rho_t}{\sum_{k\neq i}^{3}p_k}\right)-p_iC_{l_1}(\rho_i)\Bigg|\nonumber\\
&&\geq\Bigg|\Big|p_jC_{l_1}(\rho_j)-p_kC_{l_1}(\rho_k)\Big|-p_iC_{l_1}(\rho_i)\Bigg|.
\end{eqnarray}

Summing over all the $i,j,k$ in (\ref{pfth12}), we have
\begin{eqnarray}\label{pfth13}
C_{l_1}(\rho)\geq\frac{1}{3}\sum_{i\neq \{j,k\},j<k}\Bigg|\Big|p_jC_{l_1}(\rho_j)-p_kC_{l_1}(\rho_k)\Big|-p_iC_{l_1}(\rho_i)\Bigg|,
\end{eqnarray}
where $G_k^{(2)}=\Big|p_iC_{l_1}(\rho_i)-p_jC_{l_1}(\rho_j)\Big|$, $i< j$, $k\neq \{i,j\}$, $i,j,k\in\{1,2,3\}$. Thus (\ref{pfth13}) can be rewritten as $C_{l_1}(\rho)\geq\frac{1}{3}\sum_{k=1}^3\big|G_k^{(2)}-p_kC_{l_1}(\rho_k)\big|$.

(\ref{th1}) reduces to (\ref{pfth13}) for $n=3$. Now suppose That Theorem 1 holds for $n=m$. Consider the case of $n=m+1$, $\rho=\sum_{i=1}^{m+1}p_i\rho_i$, we have
\begin{eqnarray}\label{pfth14}
C_{l_1}(\rho)&&=C_{l_1}\left(\sum_{i\neq k}^{m+1}p_i\rho_i+p_k\rho_k\right)\nonumber\\
&&\geq\bigg|\sum_{i\neq k}^{m+1}p_i C_{l_1}\left(\frac{\sum_{j\neq k}^{m+1}p_j\rho_j}{\sum_{t\neq k}^{m+1}p_t}\right)-p_kC_{l_1}(\rho_k)\bigg|\nonumber\\
&&\geq \bigg|G_k^{(m)}-p_kC_{l_1}(\rho_k)\bigg|.
\end{eqnarray}
Summing over all the $k$ in (\ref{pfth14}), we have
\begin{eqnarray}\label{pfth15}
C_{l_1}(\rho)&&\geq\frac{1}{m+1}\sum_{k=1}^{m+1} \bigg|G_k^{(m)}-p_kC_{l_1}(\rho_k)\bigg|.
\end{eqnarray}
Therefore, the left inequality of (\ref{th1}) is proved. The right inequality in (\ref{th1}) is obvious due to the convex property of the $l_1$-norm coherence. $\Box$

Note that here the decomposed states $\rho_1,\cdots,\rho_n$ can be either pure or mixed states.
For pure state decompositions, we have

{[\bf Theorem 2]}.
If $\rho$ has a linearly independent pure state decomposition $|\psi_1\ra,~|\psi_2\ra,\cdots,|\psi_n\ra$, $\rho=\sum_{i=1}^np_i|\psi_i\ra\la\psi_i|$, then the convex-roof $l_1$-norm coherence
of multiqubit state $\rho$ satisfies the following triangle-like inequality:
\begin{eqnarray}\label{th2}
   \frac{1}{n}\sum_{k=1}^n\bigg|\tilde {G}_k^{(n-1)}-p_k\tilde {C_{l_1}}(|\psi_k\ra)\bigg|\leq \tilde {C_{l_1}}(\rho)\leq\sum_i p_i\tilde {C_{l_1}}(|\psi_i\ra),
  \end{eqnarray}
where $\tilde {C_{l_1}}(|\psi_i\ra)={C_{l_1}}(|\psi_i\ra)$ is the convex-roof $l_1$-norm coherence of
pure states $|\psi_i\ra~i=1,2,\cdots,n$, $\tilde {G}_k^{(n-1)}$, $k=1,2,\cdots,n$, are the low bounds of $\sum_{i\neq k}^{n}p_i \tilde {C_{l_1}}\left(\frac{\sum_{j\neq k}^{n}p_j\rho_j}{\sum_{t\neq k}^{n}p_t}\right)$.

{\sf [Proof].}
The second inequality of (\ref{th2}) is due to that the convex-roof $l_1$-norm coherence $\tilde {C_{l_1}}(\rho)$ is a sum of
the minimal decomposition of $\rho$, while $\sum_i p_iC_{l_1}(|\psi_i\ra)$ can be regarded as a sum of a general decomposition of $\rho$.

Assume that $\rho=\sum_kp_k|\psi_k\ra\la\psi_k|$ is the optimal decomposition of $\tilde {C_{l_1}}(\rho)$. We have
\begin{eqnarray}\label{pfth22}
\tilde {C_{l_1}}(\rho)&&=\sum_k p_kC_{l_1}(|\psi_k\ra)\nonumber\\
&&=\sum_k p_k\sum_{i\ne j}\big|\la i|\psi_k\ra\la\psi_k|j\ra\big|\nonumber\\
&&\geq \sum_{i\ne j}\big|\la i|\sum_kp_k|\psi_k\ra\la\psi_k|j\ra\big|\nonumber\\
&&=\sum_{i\ne j}|\rho_{ij}|=C_{l_1}(\rho).
\end{eqnarray}
From Theorem 1, we obtain (\ref{th2}), which completes the proof. $\Box$

{\it Remark 1}. Theorem 1 is the main result in \cite{dy}, where $n=2$.  There is a potential question of whether other quantum measures also satisfy triangle inequalities. One may also consider other measures of coherence, and entanglement measures such as relative entropy and the entanglement of formation. However, these measures are usually difficult to calculate for given states in general. Hence, the triangle-like inequalities associated with these measures are still open problems.

\section{triangle-like inequalities in entanglement negativity}
Concurrence is a well-known measure of entanglement \cite{rpmk,og,kfb,pvc,ppm}. For a general bipartite pure state $|\psi\rangle_{AB}$ in $H_A\otimes H_B$, the concurrence is defined by \cite{AU,pvc,SA},
$C(|\psi\rangle_{AB})=\sqrt{{2\left[1-\mathrm{Tr}(\rho_A^2)\right]}},$
with $\rho_A$ the reduced density matrix by tracing over the subsystem $B$, $\rho_A=\mathrm{Tr}_B(|\psi\rangle_{AB}\langle\psi|)$.
The concurrence for a bipartite mixed state $\rho_{AB}$ is defined by the convex-roof extension
$C(\rho_{AB})=\min_{\{p_i,|\psi_i\rangle\}}\sum_ip_iC(|\psi_i\rangle)$, where the minimum is taken over all possible decompositions of $\rho_{AB}=\sum_ip_i|\psi_i\rangle\langle\psi_i|$, with $p_i\geq0$ and $\sum_ip_i=1$ and $|\psi_i\rangle\in H_A\otimes H_B$.

Another well-known measure of bipartite entanglement is the negativity. Given a bipartite state $\rho_{AB}$ in $H_A\otimes H_B$, the negativity is defined by \cite{GRF},
$N(\rho_{AB})=(||\rho_{AB}^{T_A}||-1)/2$,
where $\rho_{AB}^{T_A}$ is the partial transpose of $\rho_{AB}$ with respect to the subsystem $A$, $||X||$ denotes the trace norm of $X$, $||X||=\mathrm{Tr}\sqrt{XX^\dag}$.
Negativity is a computable measure of entanglement, and is a convex function of $\rho_{AB}$. It vanishes if and only if $\rho_{AB}$ is separable for the $2\otimes2$ and $2\otimes3$ systems \cite{MPR}. For simplicity, we use the following definition of negativity, $N(\rho_{AB})=||\rho_{AB}^{T_A}||-1$.
For a mixed state $\rho_{AB}$, the convex-roof extended negativity (CREN) is defined by $$N_c(\rho_{AB})=\mathrm{min}\sum_ip_iN(|\psi_i\rangle_{AB}),$$ where the minimum is taken over all possible pure state decompositions $\{p_i,~|\psi_i\rangle_{AB}\}$ of $\rho_{AB}$. CREN gives a perfect discrimination of positive partial transposed bound entangled states and separable states in any bipartite quantum systems \cite{PH,WJM}.

For any bipartite pure state $|\psi\rangle_{AB}$ in a $d\otimes d$ quantum system with Schmidt rank 2,
$|\psi\rangle_{AB}=\sqrt{\lambda_0}|00\rangle+\sqrt{\lambda_1}|11\rangle$,
one has
$N(|\psi\rangle_{AB})=\parallel|\psi\rangle\langle\psi|^{T_B}\parallel-1=2\sqrt{\lambda_0\lambda_1}
=\sqrt{2(1-\mathrm{Tr}\rho_A^2)}=C(|\psi\rangle_{AB})$. Namely, the negativity is equivalent to the concurrence for any pure state with Schmidt rank 2. Consequently it follows that for any two-qubit mixed state $\rho_{AB}=\sum p_i|\psi_i\rangle_{AB}\langle\psi_i|$,
\begin{eqnarray}\label{ne}
 N_c(\rho_{AB})=\mathrm{min}\sum_ip_iN(|\psi_i\rangle_{AB})=\mathrm{min}\sum_ip_iC(|\psi_i\rangle_{AB})=C(\rho_{AB}),
\end{eqnarray}
For a general bipartite pure state $\rho_{AB}=|\psi\ra\la\psi|$, the concurrence can be written as \cite{sja}, $C(|\psi\ra)=2\sqrt{\sum_{i< j}^{d_1(d_1-1)/2}\sum_{k< l}^{d_1(d_1-1)/2}\Big|a_{ik}a_{jl}-a_{il}a_{jk}\Big|^2}$, with $|\psi\ra=\sum_{i=1}^{d_1}\sum_{j=1}^{d_2}a_{ij}|ij\ra$. It is easy to obtain $p_iC(|\psi_i\ra)=C(\sqrt{p_i}|\psi_i\ra),~i=1,2$. From the result of Theorem 3 in \cite{dy}, one has,
\begin{eqnarray}\label{con}
\Big|p_1C(|\psi_1\ra)-p_2C(|\psi_2\ra)\Big|\leq C(\rho)\leq p_1C(|\psi_1\ra)+p_2C(|\psi_2\ra),
\end{eqnarray}
where $\rho=p_1|\psi_1\ra\la\psi_1|+p_2|\psi_2\ra\la\psi_2|$ is a  two-qubit mixed state.

Combining (\ref{ne}) and (\ref{con}), we have
\begin{eqnarray}\label{nc}
\Big|p_1N_c(|\psi_1\ra)-p_2N_c(|\psi_2\ra)\Big|\leq N_c(\rho)\leq p_1N_c(|\psi_1\ra)+p_2N_c(|\psi_2\ra),
\end{eqnarray}
and with a similiar method of Theorem 1, we generalize (\ref{nc}) to multiqubit states.

{[\bf Theorem 3]}.
If a multiqubit state $\rho$ can be expressed as a convex
combination of $n$ ($n\geq 2$) states $\rho=\sum_{i=1}^np_i\rho_i$, with $\rho_i=|\psi_i\ra\la\psi_i|,~i=1,2,\cdots,n$, then the negativity $N_c(\rho)$ satisfies the triangle-like inequality:
\begin{eqnarray}\label{th3}
   \frac{1}{n}\sum_{k=1}^n\bigg|H_k^{(n-1)}-p_kN_c(\rho_k)\bigg|\leq N_c(\rho)\leq\sum_i p_iN_c(\rho_i),
  \end{eqnarray}
where $H_k^{(n-1)}$, $k=1,2,\cdots,n$, are the low bounds of $\sum_{i\neq k}^{n}p_i N_c\left(\frac{\sum_{j\neq k}^{n}p_j\rho_j}{\sum_{t\neq k}^{n}p_t}\right)$, $H_1^{(1)}=p_2N_c(\rho_2)$, $H_2^{(1)}=p_1N_c(\rho_1)$.

{\sf [Proof].}
First, we consider the case of $n=3$, $\rho=\sum_{i=1}^3p_i\rho_i$. From Eq. (\ref{nc}), we get

\begin{eqnarray}\label{pfth32}
N_c(\rho)&&\geq \Bigg|\sum_{j\neq i}^{3}p_jN_c\left(\frac{\sum_{t\neq i}^{3}p_t\rho_t}{\sum_{k\neq i}^{3}p_k}\right)-p_iN_c(\rho_i)\Bigg|\nonumber\\
&&\geq\Bigg|\Big|p_jN_c(\rho_j)-p_kN_c(\rho_k)\Big|-p_iN_c(\rho_i)\Bigg|.
\end{eqnarray}

Summing over all the $i,j,k$ in (\ref{pfth32}), we have
\begin{eqnarray}\label{pfth33}
N_c(\rho)\geq\frac{1}{3}\sum_{i\neq \{j,k\},j<k}\Bigg|\Big|p_jN_c(\rho_j)-p_kN_c(\rho_k)\Big|-p_iN_c(\rho_i)\Bigg|,
\end{eqnarray}
where $H_k^{(2)}=\Big|p_iN_c(\rho_i)-p_jN_c(\rho_j)\Big|$, $i\leq j$, $k\neq \{i,j\}$, $i,j,k\in\{1,2,3\}$. Then (\ref{pfth33}) can be rewritten as $N_c(\rho)\geq\frac{1}{3}\sum_{k=1}^3\big|H_k^{(2)}-p_kN_c(\rho_k)\big|$.

(\ref{th3}) reduces to (\ref{pfth33}) for $n=3$. Suppose that Theorem 3 holds for $n=m$. Consider the state $\rho=\sum_{i=1}^{m+1}p_i\rho_i$, we have
\begin{eqnarray}\label{pfth34}
N_c(\rho)&&=N_c\left(\sum_{i\neq k}^{m+1}p_i\rho_i+p_k\rho_k\right)\nonumber\\
&&\geq\bigg|\sum_{i\neq k}^{m+1}p_i N_c\left(\frac{\sum_{j\neq k}^{m+1}p_j\rho_j}{\sum_{t\neq k}^{m+1}p_t}\right)-p_kN_c(\rho_k)\bigg|\nonumber\\
&&\geq \bigg|H_k^{(m)}-p_kN_c(\rho_k)\bigg|.
\end{eqnarray}
Summing over all the $k$ in (\ref{pfth34}), we have
\begin{eqnarray}\label{pfth15}
N_c(\rho)&&\geq\frac{1}{m+1}\sum_{k=1}^{m+1} \bigg|H_k^{(m)}-p_kN_c(\rho_k)\bigg|.
\end{eqnarray}
Therefore, the left inequality of (\ref{th3}) is proved. The right inequality in (\ref{th3}) is obvious due to the convex property of the negativity. $\Box$

{\it Remark 2}. Theorem 3 can be used to detect the entanglement of a multiqubit mixed state. If the product of the probability distribution of $|\psi_i\ra$ and its corresponding negativity is different from that of $|\psi_j\ra$, $\rho$ must be entangled. If all $|\psi_i\ra$, $i=1,2,\cdots,n$, are partial transpose positive states, $\rho$ must be separable.  In the following example, one can find that for $p=\frac{1}{3}$, the left hand side of Theorem 3 is 0, but the state $\rho$ is entangled.

{\it Example.} Consider a two-qubit mixed state $\rho=p|\psi_1\ra\la\psi_1|+(1-p)|\psi_2\ra\la\psi_2|$, where $|\psi_1\ra=\sqrt{\frac{3}{8}}(|00\ra+|11\ra)+\sqrt{\frac{1}{8}}i(|01\ra+|10\ra)$, $|\psi_2\ra=\sqrt{\frac{3}{8}}(|00\ra+|11\ra)+\sqrt{\frac{1}{8}}(|01\ra+|10\ra)$, $i=\sqrt{-1}$. From the definition of negativity, we have $N_c(|\psi_1\ra)=1,~N_c(|\psi_2\ra)=\frac{1}{2}$ and $N_c(\rho)=\frac{1}{2}\sqrt{1+3p^2}$. The relations among $\Big|pN_c(|\psi_1\ra)-(1-p)N_c(|\psi_2\ra)\Big|$, $N_c(\rho)$ and $pN_c(|\psi_1\ra)+(1-p)N_c(|\psi_2\ra)$ are shown in Fig. 1.

\begin{figure}
  \centering
  \includegraphics[width=11cm]{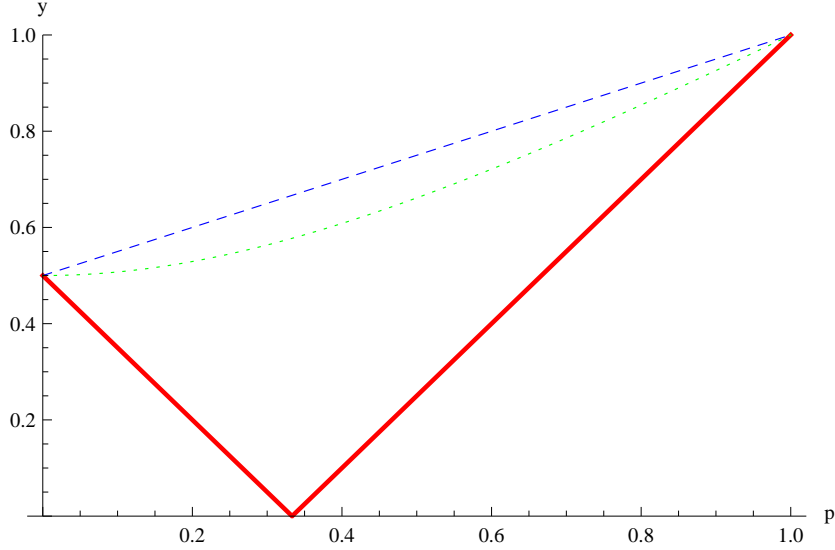}\\
  \caption{Solid (red) line denotes the value of $\Big|pN_c(|\psi_1\ra)-(1-p)N_c(|\psi_2\ra)\Big|$ for given $p$, dashed (blue) line for $pN_c(|\psi_1\ra)+(1-p)N_c(|\psi_2\ra)$, and dotted (green) line for $N_c(\rho)$.}
\end{figure}

In fact, the bounds of $N_c(\rho)$ in (\ref{th3}) depends on the pure state decompositions of $\rho$.
Consider a new decomposition $\rho=p|\psi_1'\ra\la\psi_1'|+(1-p)|\psi_2'\ra\la\psi_2'|$,
where $(\sqrt{p}|\psi_1'\ra,\sqrt{1-p}|\psi_2'\ra)^T=U(\sqrt{p}|\psi_1\ra,\sqrt{1-p}|\psi_2\ra)^T$,
$$
U=\left(
\begin{array}{cc}
 \cos{\alpha}e^{i\beta} & \sin{\alpha}e^{i\gamma}   \\
 -\sin{\alpha} e^{-i\gamma} & \cos{\alpha} e^{-i\beta}
\end{array}
\right).
$$
The relation still holds,
$$
\Big|p_1N_c(|\psi_1'\ra)-p_2N_c(|\psi_2'\ra)\Big|\leq N_c(\rho)\leq p_1N_c(|\psi_1'\ra)+p_2N_c(|\psi_2'\ra).
$$
In particular, under all possible unitary transformations $U$, the largest upper bound of
$N_c(\rho)$ is just the convex-roof extended negativity of assistance \cite{JAB},
$N_a(\rho)=\mathrm{max}_{\{p_i,|\psi_i\ra\}}\sum_ip_iN(|\psi_i\rangle)$.

\section{conclusion}
We have provided a general triangle-like inequality for any $n$ pure states combinations of $\rho$, based on the $l_1$-norm coherence measure. Furthermore, we have presented the triangle-like inequality satisfied by the convex-roof extended negativity for any states of rank 2, giving a positive answer to the conjecture raised in \cite{dy}. By a detailed example of a two-qubit state, the relations characterized by the triangle-like inequalities have been explicitly displayed.
These results may highlight the further investigations on quantum coherence, quantum entanglement and even other quantum correlations related to steerability and non-locality.

\bigskip
\noindent{\bf Acknowledgments}\, \, The authors would like to thank the anonymous referees for their valuable comments which helped to improve the results of the manuscript. This work is supported by the NSF of China under Grant No. 11675113 and NSF of Beijing under No. KZ201810028042.


\begin{thebibliography}{99}
\bibitem{tmm} T. Baumgratz, M. Cramer, and M. B. Plenio, Quantifying Coherence, Phys. Rev. Lett. 113, 140401 (2014).

\bibitem{auhm} A. Streltsov, U. Singh, H. S. Dhar, M. N. Bera, and G. Adesso, Measuring Quantum Coherence with Entanglement, Phys. Rev. Lett. 115, 020403 (2015).
\bibitem{dg} D. Girolami, Observable Measure of Quantum Coherence in Finite Dimensional Systems, Phys. Rev. Lett. 113, 170401 (2014).
\bibitem{yxlc} Y. Yao, X. Xiao, L. Ge, and C. P. Sun, Quantum coherence in multipartite systems, Phys. Rev. A 92, 022112 (2015).
\bibitem{chs} C. S. Yu, and H. S. Song, Bipartite concurrence and localized coherence, Phys. Rev. A 80, 022324 (2009).

\bibitem{aer} A. E. Rastegin, Quantum-coherence quantifiers based on the Tsallis relative $\alpha$ entropies, Phys. Rev. A 93, 032136 (2016).
\bibitem{jbdv} J. Ma, B. Yadin, D. Girolami, V. Vedral, and M. Gu, Converting Coherence to Quantum Correlations, Phys. Rev. Lett. 116, 160407 (2016).

\bibitem{ad} A. Winter, and D. Yang, Operational Resource Theory of Coherence, Phys. Rev. Lett. 116, 120404 (2016).

\bibitem{spm} S. Rana, P. Parashar, and M. Lewenstein, Trace-distance measure of coherence, Phys. Rev. A 93, 012110 (2016).


\bibitem{easm} E. Chitambar, A. Streltsov, S. Rana, M. N. Bera, G. Adesso, and M. Lewenstein, Assisted Distillation of Quantum Coherence, Phys. Rev. Lett. 116, 070402 (2016).



\bibitem{csb} C. S. Yu, S. R. Yang, and B. Q. Guo, Total quantum coherence and its applications, Quant. Inf. Proc. 15, 3773 (2016).
\bibitem{eg} E. Chitambar, and G. Gour, Critical Examination of Incoherent Operations and a Physically Consistent Resource Theory of Quantum Coherence, Phys. Rev. Lett. 117, 030401 (2016).


\bibitem{ctmm} C. Napoli, T. R. Bromley, M. Cianciaruso, M. Piani, N. Johnston, and G. Adesso, Robustness of Coherence: An Operational and Observable Measure of Quantum Coherence, Phys. Rev. Lett. 116, 150502 (2016).
\bibitem{em} E. Chitambar, and M. H. Hsieh, Relating the Resource Theories of Entanglement and Quantum Coherence, Phys. Rev. Lett. 117, 020402 (2016).
\bibitem{ir} I. Marvian, and R. W. Spekkens, Modes of asymmetry: The application of harmonic analysis to symmetric quantum dynamics and quantum reference frames, Phys. Rev. A 90, 062110 (2014).
\bibitem{irp} I. Marvian, R. W. Spekkens, and P. Zanardi, Quantum speed limits, coherence, and asymmetry, Phys. Rev. A 93, 052331 (2016).

\bibitem{ula} U. Singh, L. Zhang, and A. K. Pati, Average coherence and its typicality for random pure states, Phys. Rev. A 93, 032125 (2016).

\bibitem{mmtc} M. Piani, M. Cianciaruso, T. R. Bromley, C. Napoli, N. Johnston, and G. Adesso, Robustness of asymmetry and coherence of quantum states, Phys. Rev. A 93, 042107 (2016).
\bibitem{cmst} C. Radhakrishnan, M. Parthasarathy, S. Jambulingam, and T. Byrnes, Distribution of Quantum Coherence in Multipartite Systems, Phys. Rev. Lett. 116, 150504 (2016).
\bibitem{bukf} K. F. Bu, U. Singh, S. M. Fei, A. K. Pati, J. D. Wu, Maximum Relative Entropy of Coherence: An Operational Coherence Measure, Phys. Rev. Lett. 119, 150405 (2017).



\bibitem{yzcm} X. Yuan, H. Zhou, Z. Cao, X. Ma, Intrinsic randomness as a measure of quantum coherence, Phys. Rev. A 92, 022124 (2015).

\bibitem{gt} X. Qi, T. Gao, and F. L. Yan, Measuring coherence with entanglement concurrence, J. Phys. A: Math. Theor. 50, 285301 (2017).

\bibitem{dbq} S. Du, S. Bai, X. Qi, Coherence measures and optimal conversion for coherent states, Quantum Inf. Comput. 15, 1307 (2015).


\bibitem{bosp}M. N. Bera, T. Qureshi, M. A. Siddiqui, A. K. Pati, Duality of quantum coherence and path distinguishability, Phys. Rev. A 92, 012118 (2015).
\bibitem{bbch} E. Bagan, J. A. Bergou, S. S. Cottrell, M. Hillery, Relations between Coherence and Path Information, Phys. Rev. Lett. 116, 160406 (2016).

\bibitem{bukf1} K. F. Bu, L. Li, J. D. Wu, S. M. Fei, Duality relation between coherence and path information in the presence of quantum memory, J. Phys. A, 085304 (2018).

\bibitem{ch} S. Cheng, M. J. W. Hall, Complementarity relations for quantum coherence, Phys. Rev. A 92, 042101 (2015).
\bibitem{sbdp} U. Singh, M. N. Bera, H. S. Dhar, A. K. Pati, Maximally coherent mixed states: Complementarity between maximal coherence and mixedness, Phys. Rev. A 91, 052115 (2015).


\bibitem{dy} Y. Dai, W. You, Y. Dong, C. Zhang. Triangle inequalities in coherence measures and entanglement concurrence, Phys. Rev. A 96, 062308 (2017).


\bibitem{rpmk} R. Horodecki, P. Horodecki, M. Horodecki, and K. Horodecki, Quantum entanglement, Rev. Mod. Phys. 81, 865 (2009).
\bibitem{og} O. G$\ddot{u}$hne and G. T$\acute{o}$th, Entanglement detection, Phys. Rep. 474, 1 (2009).
\bibitem{kfb} K. Audenaert, F. Verstraete, and B. De Moor, Variational characterizations of separability and entanglement of formation, Phys. Rev. A 64, 052304 (2001).
\bibitem{pvc} P. Rungta, V. Buzek, C. M. Caves, M. Hillery, and G. J. Milburn, Universal state inversion and concurrence in arbitrary dimensions, Phys. Rev. A 64, 042315 (2001).
\bibitem{ppm} P. Badziag, P. Deuar, M. Horodecki, P. Horodecki, and R. Horodecki, Concurrence in arbitrary dimensions, J. Mod. Opt. 49, 1289 (2002).

\bibitem{AU} A. Uhlmann, Fidelity and concurrence of conjugated states, Phys. Rev. A 62, 032307 (2000).

\bibitem{SA} S. Albeverio and S. M. Fei, A note on invariants and entanglements, J. Opt. B: Quantum Semiclass Opt. 3, 223 (2001).	

\bibitem{GRF} G. Vidal, and R. F. Werner, Computable measure of entanglement, Phys. Rev. A. 65, 032314 (2002).
\bibitem{MPR} M. Horodecki, P. Horodecki, and R. Horodecki, Mixed-State Entanglement and Distillation: Is there a “Bound” Entanglement in Nature? Phys. Rev. Lett. 80, 5239 (1998).	
\bibitem{PH} P. Horodeki, Separability criterion and inseparable mixed states with positive partial transposition, Phys. Lett. A. 232, 333 (1997).
\bibitem{WJM} W. Dur, J. I. Cirac, M. Lewenstein, and D. Bru${\ss}$. Distillability and partial transposition in bipartite systems. Phys. Rev. A. 61, 062313 (2000).	


\bibitem{sja} S. J. Akhtarshenas, Concurrence vectors in arbitrary multipartite quantum systems, J. Phys. A 38, 6777 (2005).

\bibitem{JAB} J. S. Kim, A. Das, and B. S. Sanders. Entanglement monogamy of multipartite higher-dimensional quantum systems using convex-roof extended negativity. Phys. Rev. A. 79, 012329 (2009).	
\end{thebibliography}
\end{document}